# Room temperature soft ferromagnetism in the nanocrystalline form of $YCo_2$ – a well-known bulk Pauli paramagnet

*S. Narayana Jammalamadaka and E.V. Sampathkumaran[*]*
Tata Institute of Fundamental Research, Homi Bhabha Road, Colaba, Mumbai – 400005, India

*and*

*V. Satya Narayana Murthy and G. Markandeyulu*
Department of Physics, Indian Institute of Technology Madras, Chennai - 600036, India.

The Laves phase compound $YCo_2$ is a well-known exchange-enhanced Pauli paramagnet. We report here that, in the nanocrystalline form, this compound interestingly is an itinerant ferromagnet at room temperature with a low coercive field. The magnitude of the saturation moment (about $1\mu_B$ per formula unit) is large enough to infer that the ferromagnetism is not a surface phenomenon in these nanocrystallites. Since these ferromagnetic nanocrystallites are easy to synthesize with a stable form in air, one can explore applications, particularly where the hysteresis is a disadvantage.

Synthesis of materials with ferromagnetism, particularly in nanoform, at room temperatures continues to be of great interest considering application potential. The magnetic-field-induced phase transition from paramagnetism to ferromagnetic state by the application of a magnetic field (H), called 'itinerant electron metamagnetism (IEM)' has been a subject of great interest and it is worthwhile to explore whether such materials can be driven ferromagnetic at room temperature in some way in the absence of a magnetic field. The topic of IEM was first addressed theoretically by Wohlfarth and Rhodes as early as 1962 [1]. Among the few candidates that dominated attention initially, $YCo_2$, crystallizing in cubic Laves phase, is a pronounced candidate. This compound is an example for exchange-enhanced paramagnet [2] and a magnetic-field-induced IEM was first reported for an application of a magnetic field of about 700 kOe [3], which generated a lot of activity [4]. There is a recent interest on this compound with respect to surface magnetism [5, 6]. In this article, we report that this Pauli paramagnet in fact is a ferromagnet, apparently not restricted to surface, in the nanocrystalline form. The transformation of a bulk Pauli paramagnet to the ferromagnetic state in the nanocrystalline form had not been reported previously in the literature. These ferromagnetic nanocrystals are found to be stable in air thereby enabling potential applications.

The polycrystalline $YCo_2$ in the ingot form (labeled *A*) was prepared by repeated melting of required amounts of high purity Y (99.9%) and Co (99.99%) in an arc furnace in an atmosphere of argon. The sample thus obtained was characterized by x-ray diffraction (XRD) (Cu $K_α$) to be single phase (within the detection limit of 1%). In addition, the back-scattered scanning electron microscopic (SEM) pictures (obtained with JEOL JSM 840A) confirm single phase nature and homogeneity and composition was further ensured by energy dispersive x-ray (EDX) analysis. The ingot thus prepared was used to prepare the nano-sized specimens. The material was milled in a planetary ball mill (Fritsch pulverisette-7 premium line) operating at a speed of 800 rpm in a medium of toluene. Tungsten carbide vials and balls of 5 mm diameter were used with a balls-to-material mass ratio of 20:1. The specimens employed for investigations were the ones milled for 15 (labeled *B*) and 45 minutes (labeled *C*). While the specimens were stored in toluene medium all the time, XRD and SEM/EDX analysis at frequent intervals of time after completion of milling revealed that these powders are stable in air over a period of several days. It appears that there is an initial oxidation for *C* resulting in the formation of a small amount of $Y_2O_3$ (marked by asterisk in figure 1) immediately after exposure to air and (it is fascinating that) further oxidation is arrested or very slow presumably due to natural surface coating by this oxide layer. It appears that toluene also plays a role for this stability in air, as the specimen heated in vacuum, say at 473 K for 1 hr, resulting in a possible loss of toluene cap, has been found to degrade faster on exposure to air. No other phase containing Y and Co could be detected in x-ray diffraction and SEM/EDX data following milling. SEM, transmission electron microscope (TEM) (Tecnai 200kV) and Dynamical Light Scattering methods, in addition to XRD, were employed to determine the particle size. Magnetization (M) measurements (4.2 – 300 K) were carried out with the help of a commercial superconducting quantum interference device (Quantum Design, USA). A differential scanning calorimeter (NETZSCH, 200 PC) was employed to determine Curie temperature ($T_C$). About 15 milli gram was sealed in aluminium pan and the experiment was carried out in nitrogen gas atmosphere. The sample was heated from room temperature to 810 K at the rate of 10 K per minute.

The XRD diffraction patterns for few diffraction lines are shown in figure 1 for the specimens *A*, *B* and *C*. The diffraction lines (and hence lattice constant, $a = 7.223 \pm 0.002$ Å) do not shift for the milled specimens with respect to *A*. If there is a stress due to milling, one would have seen a significant shift in the position of diffraction lines, as demonstrated for amorphous nanocolumns created by heavy-ion irradiation of thin films of $YCo_2$ [Ref. 7]. There is a reduction in the intensity with decreasing particle size which is not uncommon [8] for ball-milled particles. A rough estimate of the average particle sizes can be inferred from the Scherrer's equation from the knowledge of line-width, which yields a value in the range of 50-60 nm and 40-50 nm for specimens *B* and *C* respectively. It appears that an increase in milling time reduces the particle size only marginally. Though, stress due to ball-milling can vitiate the size-determination from the diffraction pattern, it is clear that the nanoparticles obtained are still crystalline and not amorphous. At this juncture, we would like to mention that the shape and the counts of the background curve in the entire range of the diffraction pattern for **B** and *C* are found to be essentially the same as that of parent material (*A*), thereby implying that the formation of any amorphous phase can be ignored. A better idea of the nanoform on the specimen *C* has been inferred by careful SEM and TEM pictures on the particles obtained by ultrasonification in a medium of alcohol for about 30 mins. These images are shown in figure 2. The large particle in the SEM image is an agglomerate of several nanocrystallites. The TEM images clearly reveal that the particles fall in the range of about 20 to 30 nm. In addition, we have confirmed the particle sizes from dynamical light scattering method.

We now discuss the magnetization behavior (figures 3 and 4). At this juncture, we would like to state that the ingot that was employed to prepare nanomaterials was verified to reproduce a well-known broad peak in the temperature dependence of magnetic susceptibility around 200 K. In figure 3 (top), we show the M data for nanomaterials, *B* and *C*, obtained in a field of 5 kOe as a function of temperature. The first and foremost conclusion is that the values are few orders of magnitude large compared to that of the ingot. The variation up to 330 K is weak and smooth, as though a magnetic transition occurs at a higher temperature only. The values for both the specimens are rather close in the temperature range of investigation. M undergoes a dramatic increase for initial applications of H as shown in figure 3 (bottom). While there is a gradual increase (following the rise at low fields) without any clear evidence for saturation for 300K-curves, M tends to saturate beyond about 10 kOe for 35 K. The value of the saturation magnetic moment ($\mu_{sat}$) obtained from linear extrapolation of the high field data to zero field turns out to be 0.93 and 0.89 $\mu_B$ per formula unit for specimen *B* and *C* respectively at 35 K, whereas corresponding values are lower at 300 K. These values indicate absence of any dramatic effect of increasing milling time on magnetism. The value of the spontaneous moment at 35 K (300 K) is about 0.12 (0.08) and 0.2 (0.1) $\mu_B$ per formula unit for specimen *B* and *C* respectively. These results imply that the nanocystals are ferromagnetic. An important point to be stressed is that the value of $\mu_{sat}$ is not negligible, but comparable to those obtained [3] at high fields for the bulk form of the compound following IEM. This conclusively establishes that the observed ferromagnetism arises not only from the surface but also from the core of the nanocrystals. If surface alone contributes to ferromagnetism, the value of $\mu_{sat}$ is expected to be about few orders of magnitude smaller [9]. A possible surface oxide contribution to ferromagnetism (as demonstrated recently [10] for nanoform of oxides) is negligible, for instance, judged by the magnitudes of the magnetic moment (comparing in the same units) reported in Refs. 9 and 10. The present finding is noteworthy considering that, according to recent theoretical predictions [5], the magnetism in $YCo_2$ can not proceed beyond two Co layers and the bulk/core should be non-magnetic, as the sizes of the particles under study are considerably large. This naturally means that, when one deals with nanocrystals, additional factors play a role. At this juncture, it is worth recalling that the substitution of small amounts of Al for Co induces ferromagnetism at low temperatures [4] and several explanations owing to Al were proposed (see Ref. 11). (i) A shift of Fermi level due to a change in the d-electron concentration; (ii) A change in the d-bandwidth due to lattice constants change following Al-substitution, and (iii) hybridization between 3d states of Co and 3p states of Al. The present results without any doping clearly reveal that additional factors are needed to be considered. Electronic structure calculations reveal [5, 6] that the Fermi level lies near a sharply falling portion of density of states and it is possible that band narrowing in the nanoform could be responsible for the observed ferromagnetism.

We observe a very weak hysteresis at both 300K and 35 K consistent with ferromagnetism (see figure 4). The small values of coercive fields (close to 50 and 135 Oe respectively) suggest that the nanocrystalline form of this compound is a soft ferromagnet. On the other hand, much larger values of coercive fields were reported [12] for the amorphous thin films. It also appears that the amorphous films are not stable in air, requiring protection from oxidation by thick $Si_3N_4$ layers [13]. Therefore, it may not be easy to explore such amorphous films for any applications. On the other hand, the as-prepared nanocrystalline form is found to be stable in air over several days and therefore it is potentially useful for applications, particularly where hysteresis loss is to be avoided.

Finally, as shown in the inset of figure 3, there is a dip at 563 K in the differential scanning calorimetric data, for instance, for specimen *C*, attributable to magnetic transition temperature. There is a weak shoulder at a marginally lower temperature which could arise from particles from slightly different sizes. There is no evidence for any more peak at any other temperature, which establishes that the spread in particle size is negligible. No combination of Y and Co has been known to order magnetically at this temperature in the literature.

Summarizing, a well-known bulk Pauli paramagnet is driven to a ferromagnetic state at room temperature in nanocrystalline form. This work opens up an avenue for further theoretical and experimental investigations of the topic of itinerant electron magnetism in nano-sized crystalline intermetallics. Since this nanocrystalline ferromagnet is easy to synthesize in a stable form, one can explore applications.

We would like to thank Kartik K Iyer, Niharika Mohapatra and Baghyashree A Chalke for their help in some experimental part. We thank Raju V Ramanujan (Nanyang Technological University, Singapore) for bringing Ref. 8 to our attention.

*Corresponding author: sampath@mailhost.tifr.res.in

Figure 1:
(color online) X-ray diffraction patterns for molten ingot (*A*) and for the specimens of YCo$_2$ obtained by milling for 15 min (*B*) and 45 min (*C*). The weak lines attributable to Y$_2$O$_3$ due to surface oxidation following initial exposure to air are indicated by asterisks.

Figure 2:
A typical SEM (left) and TEM (right) images of the specimen *C* of YCo$_2$.

Figure 3:
(color online) (Top) Magnetization divided by temperature obtained in a field of 5 kOe in the temperature range 4.2 to 300 K for nanocrystals **B** and **C** of YCo$_2$. (Bottom) Isothermal magnetization behavior at 35 and 300 K. The lines through the data points are guides to the eyes. In the inset of top figure, the temperature dependence of heat-flow in the differential scanning calorimetric studies on nanocrystals *C* is shown.

Figure 4:
(color online) Hysteresis loops for nanocrystals, **B** and **C**, of YCo$_2$, at 35 and 300 K. The data points are joined by lines.

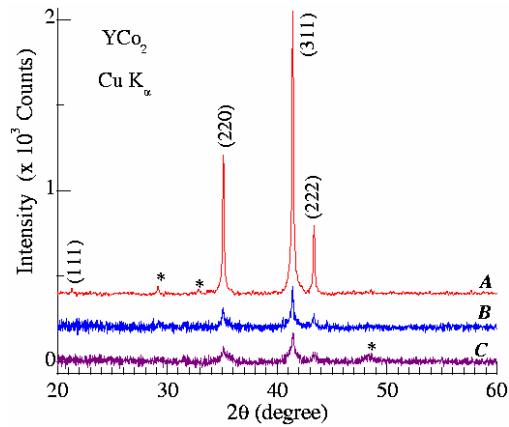

Figure 1

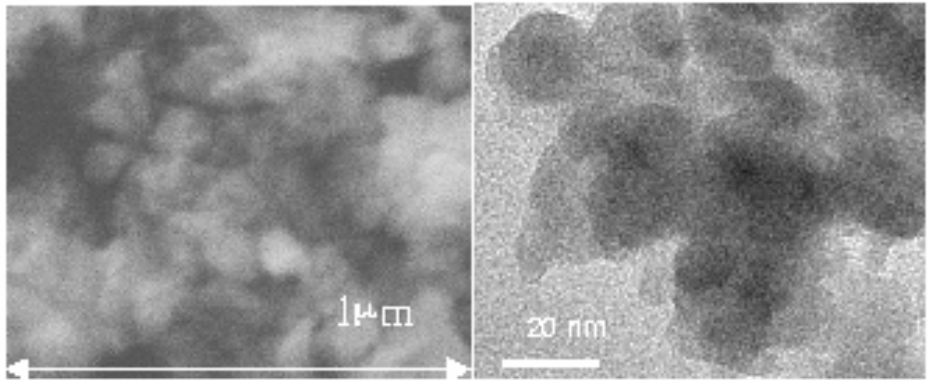

Figure 2

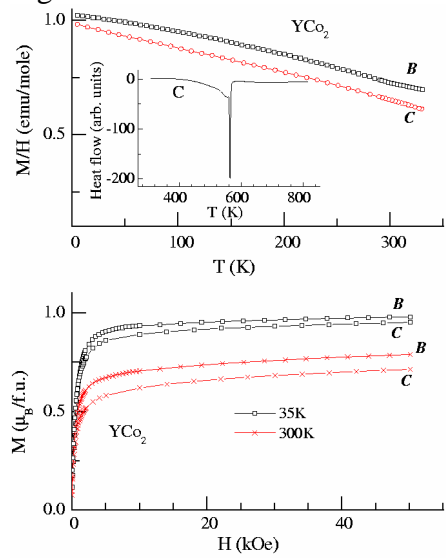

Figure 3

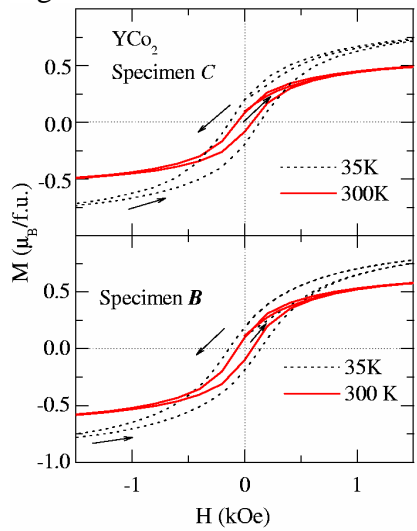

Figure 4